\documentclass{aa} \usepackage{times,float}
\usepackage{epsfig}
\usepackage{graphics}
\usepackage{amssymb}

\begin{document}

\title{On the Ly$\alpha$ emission from gamma-ray burst host galaxies: 
evidence for low metallicities
\thanks{
Based on  observations collected at the  European Southern Observatory,
in La Silla (Chile) under ESO programme ID 70.B-0233.}
\fnmsep\thanks{Based on observations made with the NASA/ESA Hubble Space 
Telescope, obtained from the data archive at the Space Telescope Institute. 
STScI is operated by the association of Universities for Research in 
Astronomy, Inc.  under the NASA contract  NAS 5-26555.}
}

\titlerunning{Ly$\alpha$ emission from GRB hosts: evidence for low 
metallicities}
\author{ 
J.P.U. Fynbo \inst{1,2} \and
P. Jakobsson \inst{2} \and
P. M\o ller \inst{3} \and
J. Hjorth \inst{2} \and
B. Thomsen \inst{1} \and
M.I. Andersen  \inst{4} \and
A.S. Fruchter \inst{5} \and
J. Gorosabel \inst{6,5} \and
S.T. Holland \inst{7} \and
C. Ledoux \inst{8} \and
H. Pedersen \inst{2} \and
J. Rhoads \inst{5} \and
M. Weidinger \inst{1} \and
R.A.M.J. Wijers \inst{9}}

\institute{  Department of Physics and Astronomy, University of Aarhus, Ny
Munkegade, DK--8000 \AA rhus C, Denmark
\and
Astronomical Observatory, University of
Copenhagen, Juliane Maries Vej 30, DK--2100 Copenhagen \O, Denmark
 \and
European Southern Observatory, Karl Schwarzschild-Strasse 2,
D-85748 Garching, Germany
\and
Astrophysikalisches Institut, D--14482 Potsdam,
Germany
\and
Space Telescope Science Institute, 3700  San Martin  Drive, Baltimore, MD
21218,   USA
\and
Instituto de Astrof\'{\i}sica de Andaluc\'{\i}a (IAA-CSIC),
P.O. Box 03004, E-18080 Granada, Spain
\and
Department of Physics, University of
Notre Dame, Notre Dame, IN 46556-5670, USA
\and
European Southern Observatory, Casilla 19001, Santiago 19,
Chile
\and
University  of Amsterdam,  Kruislaan  403, NL-1098  SJ Amsterdam,  The
Netherlands}

\mail{jfynbo@phys.au.dk}

\date{Received / Accepted }

\abstract{
We report on the results of a search for Ly$\alpha$ emission from the host 
galaxy of the $z=2.140$ GRB~011211 and other galaxies in its surrounding field.
We detect Ly$\alpha$ emission from the host as well as from six other 
galaxies in the field. The restframe equivalent width of the Ly$\alpha$ line
from the GRB~011211 host is about 21~\AA. This is the fifth detection of 
Ly$\alpha$ emission out of five possible detections from GRB host galaxies, 
strongly indicating that GRB hosts, at least at high redshifts, 
are Ly$\alpha$ emitters. This is intriguing as only $\sim$25\% of the 
Lyman-Break selected galaxies at similar redshifts have Ly$\alpha$ emission 
lines with restframe equivalent width larger than 20~\AA. Possible 
explanations are {\it i)} a preference for GRB progenitors to be metal-poor as 
expected in the collapsar model, {\it ii)} an optical afterglow selection bias 
against dusty hosts, and {\it iii)} a higher fraction of Ly$\alpha$ emitters at
the faint end of the luminosity function for high-$z$ galaxies. Of these, 
the current evidence seems to favour {\it i)}.
 } 

\maketitle

\section{Introduction}
Since 1997, the precise positional information of Gamma-Ray Burst (GRB) 
afterglows has provided a new method by which to locate and study galaxies 
in the early universe -- the host galaxies. Once an afterglow position has 
been determined detecting the host is only a matter of integrating on this 
position until the host emerges from the noise. The impact parameters of 
afterglows relative to their hosts are small enough, typically a fraction of 
an arcsec on the sky, that chance alignment is not a serious limitation 
(Bloom, Kulkarni \& Djorgovski 2002). So far this approach has led to the 
detection of a host galaxy for almost all of nearly 50 well localised GRBs. 

An important aspect of GRB 
selection compared to other selection mechanisms is that it is not flux 
limited. This is obviously the case for most other selection methods;
Lyman-Break selection (Shapley  et al. 2003 and references therein) is 
continuum flux limited and Ly$\alpha$ selection (e.g. M\o ller \& Warren 1993;
Cowie \& Hu 1998; Rhoads et al. 2000; Fynbo, M\o ller \& Thomsen 2001;
Ouchi et al. 2003; Fynbo et al. 2003) is 
line flux limited. Therefore, GRB selection allows us to probe the faint end
of the luminosity function currently inaccessible to other techniques.
GRB selection is subject to other selection mechanisms, but these are not yet 
known in detail (e.g. a relation to the occurrence of star formation). 
The precise nature of the GRB selection mechanism provides hints about 
the nature of the GRB progenitors (e.g. Woosley 1993; Paczy{\'n}ski 1998; 
Hogg \& Fruchter 1999).

Ly$\alpha$ imaging of GRB hosts is interesting as Ly$\alpha$ emitting galaxies 
(in the following we will use the acronym LEGOs -- Ly$\alpha$ Emitting 
Galaxy-building Objects, M\o ller \& Fynbo 2001) are starburst galaxies with 
little or no dust. Ly$\alpha$ imaging is therefore a probe of the star 
formation rate and of the dust content of GRB host galaxies. Both of these 
parameters are important for our 
understanding of GRB progenitors and of how the environment affects the 
propagation of afterglow emission out of host galaxies. Furthermore, 
Ly$\alpha$ narrow band imaging is an efficient way to probe if the host galaxy 
resides in an overdense environment such as a group or a proto-cluster. The 
first Ly$\alpha$ narrow band imaging of GRB host galaxies was 
presented in Fynbo et al. (2002) where we studied the fields of GRB~000301C and 
GRB~000926, both at redshift $z=2.04$. That study resulted in the detection of 
Ly$\alpha$ emission from the host of GRB~000926 and 18 additional emitters in 
the two fields. The host galaxy of GRB~000301C was too faint, $\textrm{R} 
\approx28$ (Bloom, Kulkarni \& Djorgovski 2002), to allow a detection even if 
it has a large Ly$\alpha$ equivalent width (EW). In this {\it Letter} we 
report on the results of a  
search for Ly$\alpha$ emission from the host galaxy of the $z=2.140$ GRB~011211
and other galaxies in its surrounding field. The properties of the X-ray rich 
GRB~011211 and its afterglow are discussed in Holland et al. (2002) and 
Jakobsson et al. (2003a). The redshift was measured via absorption lines in 
the spectrum of the optical afterglow to be $z=2.140$ (Fruchter et al. 2001; 
Holland et al. 2002). The host galaxy was detected with deep late time imaging 
to be a faint $\textrm{R}\approx25$ galaxy (Burud et al. 2001; Fox et al. 
2002; Jakobsson et al. 2003b).

\section{Observations and data reduction}

The observations were carried out during three nights in February
2003 at the 3.5-m New Technology Telescope on La Silla using the 
Superb Seeing Imager - 2 (SUSI2). The SUSI2 detector consists of two
2048$\times$4096 thinned, anti-reflection coated EEV CCDs with
a pixel scale of 0\farcs085. The field of GRB~011211 was imaged
in three filters: the standard B and R filters and a special
narrow-band filter manufactured by Omega Optical. The narrow-band 
filter (OO3823/59) is tuned to Ly$\alpha$ at $z=2.140$
and has a width of 59~\AA \ (corresponding to a redshift width of
$\Delta z = 0.049$ for Ly$\alpha$ or a Hubble flow depth of 4700 km
s$^{-1}$). The total integration 
times were 15 hours (OO3823/59), 3.1 hours (B-band), and 1.9 hours
(R-band). The individual exposures were bias-subtracted,
flat-field corrected and combined using standard techniques. 
The full-width-at-half-maximum (fwhm) of point sources in
the combined images are 1\farcs10 (R-band), 1\farcs11 (B-band) and
1\farcs22 (OO3823/59).

The narrow-band observations were calibrated using observations of the 
spectrophotometric standard stars LTT3218, LTT7379, and GD108 (Stone 1996).
The broad-band images were calibrated using the secondary standards from 
Jakobsson et al. (2003b) and brought onto the AB-system using the 
transformations given in Fukugita et al. (1995).

\section{Results}

We used the same methods for photometry and selection of LEGO candidates 
as those described in Fynbo et al. (2002). 

\subsection{The host galaxy}

\begin{figure}
\begin{center}
\epsfig{file=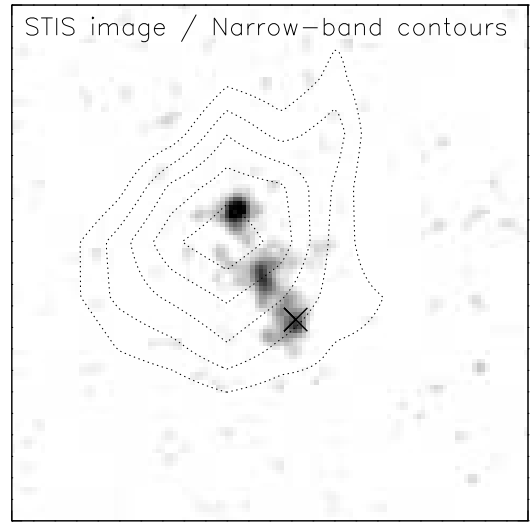,width=7.0cm}
\caption{A 3$\times$3 arcsec$^2$ section of the STIS/CL image around 
the host galaxy of GRB~011211 (Jakobsson et al. 2003b). North is up and 
East is to the left. The GRB went off in the faint, southern part of the 
object (position marked with a cross). The contours show the Ly$\alpha$ 
emission based on our narrow band observations. The Ly$\alpha$ emission 
weighted centroid of the galaxy is close to the northern knot seen in the 
STIS image.}
\label{host}
\end{center}
\end{figure}

The host galaxy of GRB~011211 is detected in all bands and is a Ly$\alpha$ 
emitter with a restframe EW of $21_{-8}^{+11}$~\AA. Deep HST/STIS images of 
the host have been reported by Fox et al. (2002) and Jakobsson et al. 
(2003b). These images show that the host is a multi-component object extending 
over almost 1 arcsec or roughly 9 kpc at $z=2.14$ (assuming $H_0$=65 km 
s$^{-1}$Mpc$^{-1}$, $\Omega_m$=0.3 and $\Omega_\Lambda$=0.7). In 
Fig.~\ref{host} we plot the contours of the Ly$\alpha$ emission on top of the 
HST/STIS image taken 59 days after the burst. The Ly$\alpha$ emission 
weighted centroid of the galaxy is close to the northern knot of the host, 
whereas the GRB occurred in the southern end of the host (Fox et al. 2002; 
Jakobsson et al. 2003b). 

\subsection{Other LEGO candidates in the field}
\begin{figure*}
\begin{center}
\epsfig{file=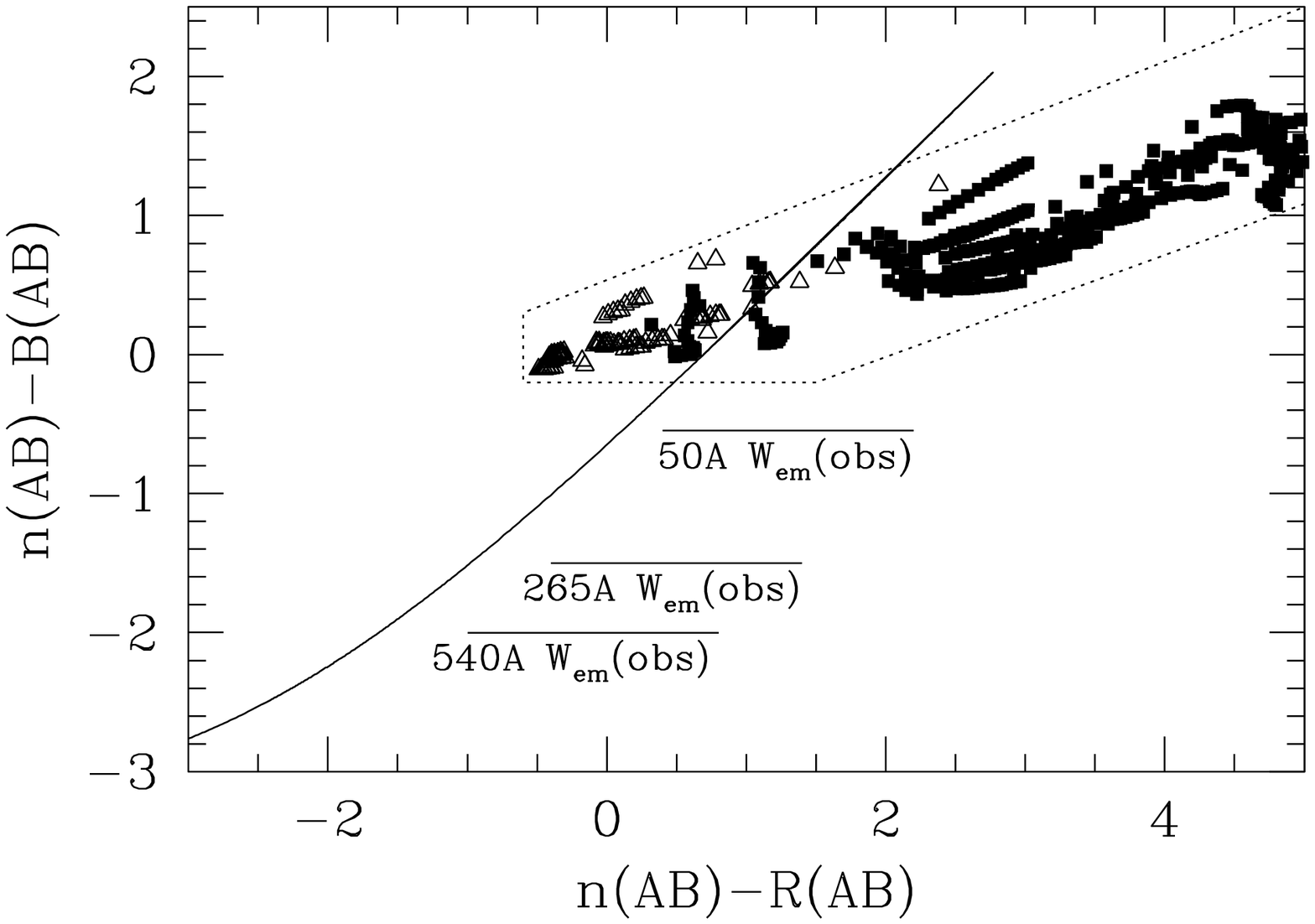,width=5.9cm}
\epsfig{file=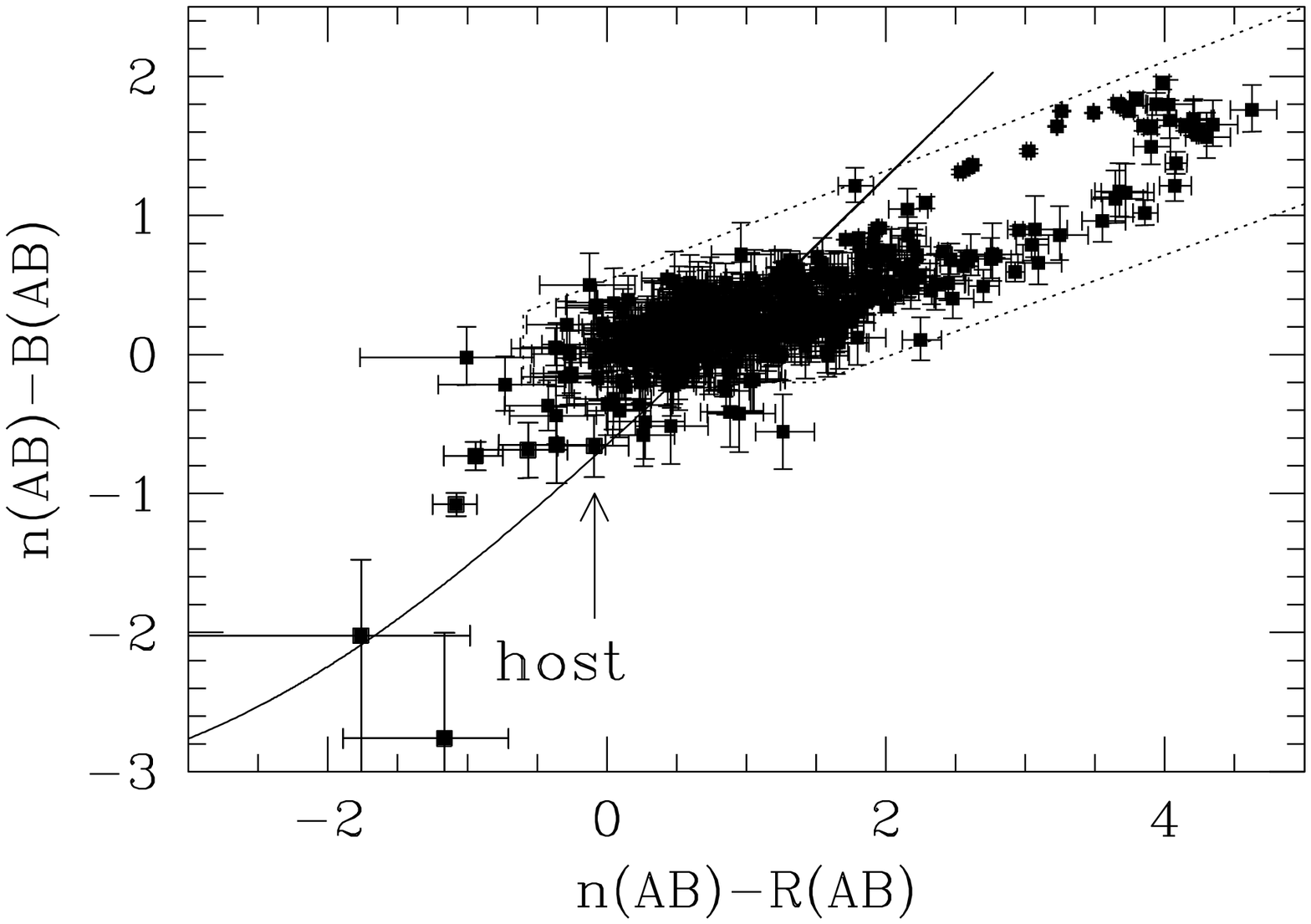,width=5.9cm}
\epsfig{file=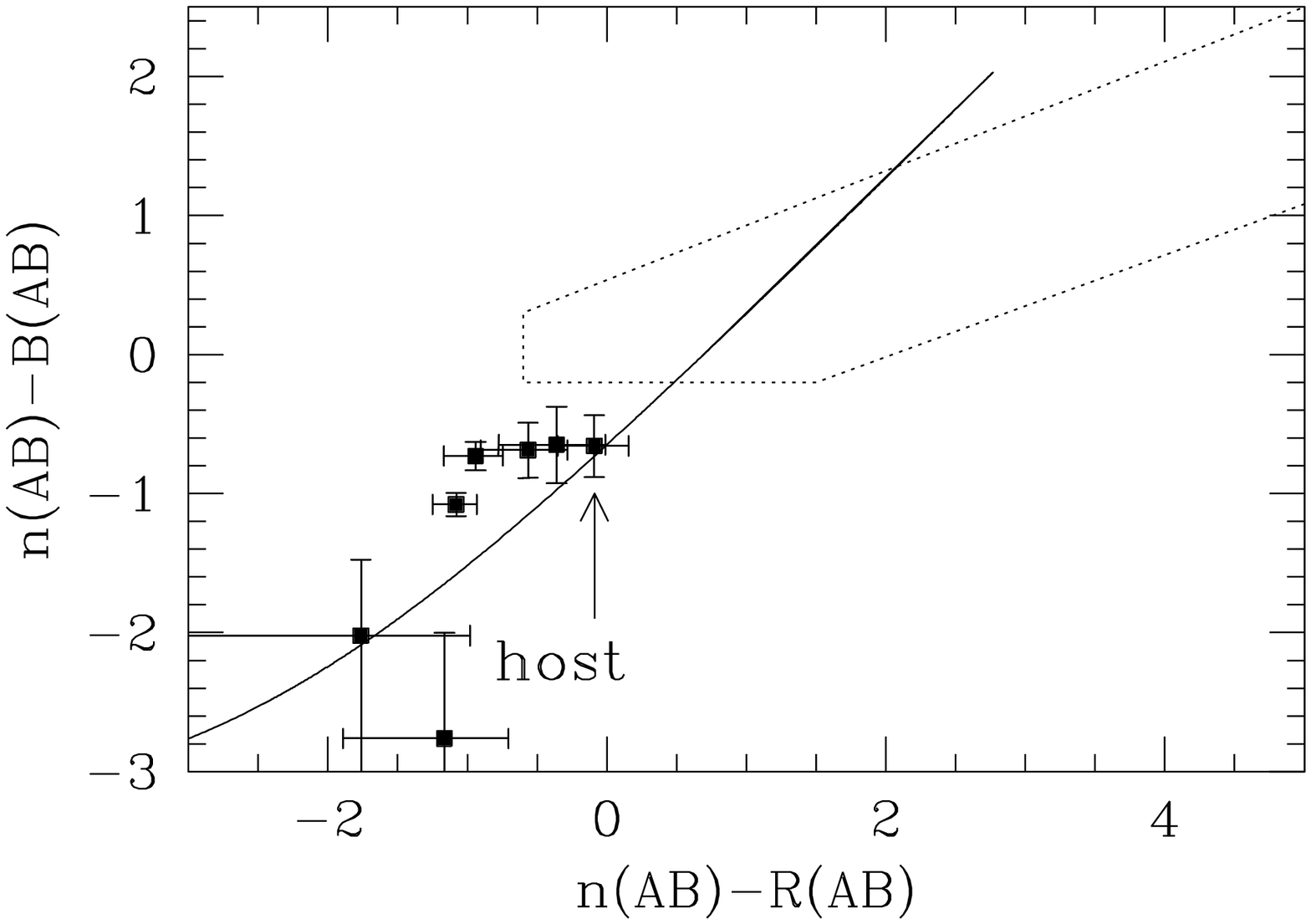,width=5.9cm}
\caption{
{\it Left panel:} Calculated colour-colour diagram based on Bruzual \& 
Charlot (1993) model galaxy spectra.
The filled squares are $0<z<0.6$ galaxies with ages from a
few to 15 Gyr and the open triangles are $0.3<z< 3.0$ galaxies
with ages from a few Myr to 1 Gyr. The dotted box contains all these
calculated galaxy colours. The full-drawn line corresponds to 
objects having the same broad-band colours, but various amounts of 
absorption (upper part) or emission (lower part) in the narrow-band filter.
{\it Middle panel:}
Colour-colour diagram for all objects in the GRB field. The
squares with error-bars indicate objects detected at S/N$>$5 in the
narrow-band image. As expected, most objects have colours consistent
with being in the dotted box. However, a number of objects, including
the GRB~011211 host, are seen in the lower left part of the diagram.
{\it Right panel:} The colours of the seven LEGO candidates which are 
selected to have n(AB)$-$B(AB)$ <-$0.60. 
}
\label{colcol}
\end{center}
\end{figure*}

\begin{figure*}
\begin{center}
\epsfig{file=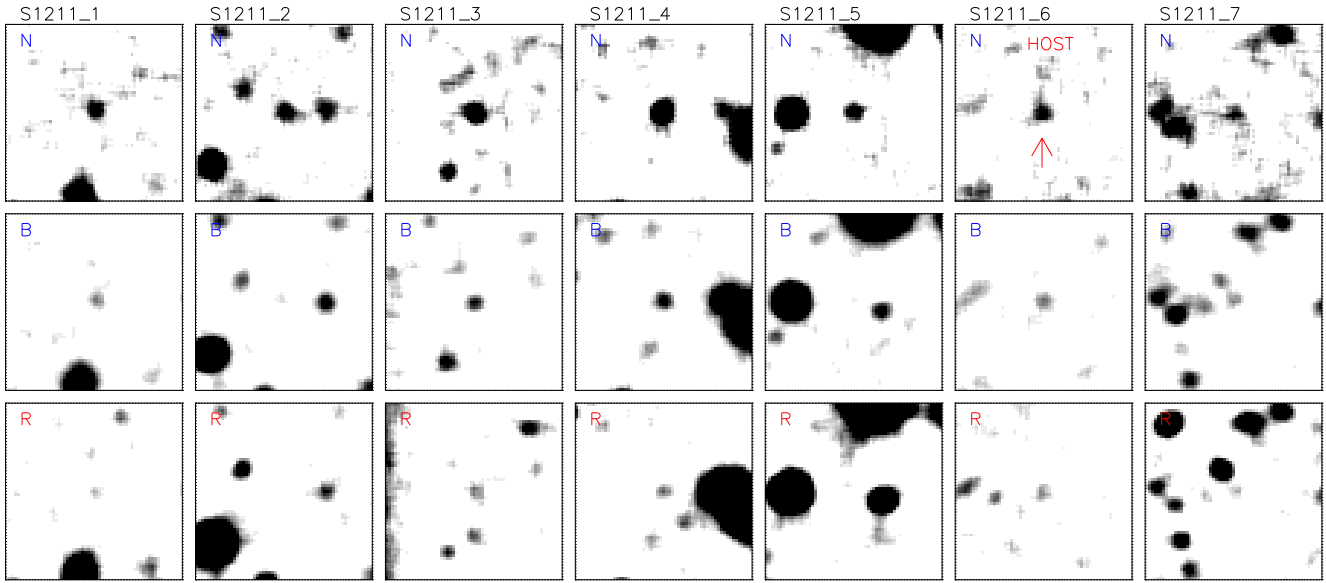,width=17.5cm}
\caption{10$\times$10 arcsec$^2$ sections around each of the seven LEGO 
candidates from the narrow band (top row), B-band (middle row) and R-band 
(bottom row). North is up and East is to the left. The host galaxy is
number six named S1211\_6.}
\label{legos}
\end{center}
\end{figure*}

In Fig.~\ref{colcol} we show the n(AB)$-$B(AB) versus n(AB)$-$R(AB) 
colour-colour 
diagram for synthetic galaxy spectra (left panel, see caption for details 
on the models) and for the detected objects in the field (middle and right
panels). The full-drawn line indicates where objects with the same broad-band 
colour and with either absorption or emission in the narrow filter will 
fall. LEGOs will fall in the lower left corner of the 
diagram (due to excess emission in the narrow-filter). In the middle panel 
we show the colour-colour diagram for all objects detected in the field. 
We select as LEGO candidates objects detected at more than 5$\sigma$ 
significance in the narrow band image and with a colour of n(AB)$-$B(AB) $< 
-$0.60. This criterion corresponds to an observed EW of about 
60~\AA, or 20~\AA \ rest for Ly$\alpha$ at $z=2.140$. The colours of the 
candidates are shown in the right panel of Fig.~\ref{colcol}. We find seven 
candidates in the 
field including the host galaxy. In the following we will refer to these as 
S1211\_1 through S1211\_7. The host galaxy does not stand out as special 
compared 
to the other candidates. It ranks sixth in brightness. Images of the candidates
are shown in Fig.~\ref{legos} and their photometric properties based on the 
total magnitudes ({\tt mag\_auto}) from SExtractor (Bertin \& Arnouts 1996) 
are given in Table~\ref{candprop}. We also derive Star Formation Rates (SFRs) 
for the LEGO candidates from the Ly$\alpha$ luminosities as described in
Fynbo et al. (2002). The seven candidates are distributed uniformly over the 
field with no obvious structure such as the $z=3.04$ filament reported by 
M\o ller \& Fynbo (2001), but this does not exclude underlying structure. 
The filter used in the present study is about three times wider than in the 
M\o ller \& Fynbo study and therefore any underlying structure would easily 
be washed out in the 2d image. 

\begin{table}
\begin{center}
\caption{Photometric properties of the seven LEGO candidates, including the host 
galaxy, in the field of GRB~011211. Upper limits are 2$\sigma$. S1211\_6 is
the host galaxy of GRB~011211.
}
\begin{tabular}{@{}lccccccc}
\hline
Object  & B(AB)  &  R(AB)  &  f(Ly$\alpha$) & SFR$_{\rm{Ly\alpha}}$ \\
        &        &         & 10$^{-17}$ &  \\
        &        &         & erg s$^{-1}$ cm$^{-2}$ & M$_{\sun}$ yr$^{-1}$ \\
\hline
S1211\_1 & 25.54$^{+0.22}_{-0.18}$ & 25.55$^{+0.55}_{-0.36}$ & 3.1$\pm$0.7 & 0.8$\pm$0.2 \\
S1211\_2 & $>$26.5 & $>$26.0 &  5.1$\pm$0.6 & 1.4$\pm$0.2 \\
S1211\_3 & 25.14$^{+0.14}_{-0.13}$ & 24.71$^{+0.21}_{-0.18}$ & 5.8$\pm$0.8 & 1.5$\pm$0.2 \\
S1211\_4 & 24.89$^{+0.10}_{-0.09}$ & 25.00$^{+0.26}_{-0.21}$ & 10.6$\pm$0.7 & 2.8$\pm$0.2 \\
S1211\_5 & $>$26.5 & $>$26.0 & 5.7$\pm$0.8 & 1.5$\pm$0.2 \\ 
S1211\_6 & 25.31$^{+0.19}_{-0.16}$ & 24.90$^{+0.30}_{-0.23}$ & 2.8$\pm$0.8 & 0.8$\pm$0.2 \\
S1211\_7 & 25.71$^{+0.19}_{-0.16}$ & 25.28$^{+0.23}_{-0.19}$ & 1.7$\pm$0.6 & 0.5$\pm$0.2 \\
\hline
\label{candprop}
\end{tabular}
\end{center}
\end{table}

\section{Discussion}

The host galaxy of GRB~011211 has been found to be a Ly$\alpha$ emitter
with a restframe EW of $21_{-8}^{+11}$~\AA. This is somewhat smaller than for 
GRB~000926 ($71_{-15}^{+20}$~\AA) and suggests the presence of more dust. 
Although 
uncertain, the UV continuum of GRB~011211 host is also redder than that 
of the GRB~000926 host. The observed B(AB)$-$R(AB) colour corresponds to 
$\beta \approx -1.2\pm0.5$, whereas Fynbo et al. (2002) found 
$\beta = -2.4\pm0.3$ and $\beta = -1.4\pm0.2$ for the two main components 
of the GRB~000926 host galaxy. Ly$\alpha$ emission has also been 
detected from the host galaxies of GRB~971214 at $z=3.42$ (Kulkarni et al. 
1998; Ahn 2000), GRB~021004 at $z=2.33$ (e.g. M\o ller et al. 2002 and 
references therein), and GRB~030323 at $z=3.37$ (Vreeswijk et al., in 
preparation).  For GRB~021004 and GRB~030323 the Ly$\alpha$ emission line is 
detected 
superimposed on the afterglow spectrum. All current evidence is consistent 
with the conjecture that the host galaxies of GRBs, at least at high 
redshifts, are Ly$\alpha$ emitters. In contrast, only $\sim$25\% and 
$\sim$33\% of the Lyman-Break selected galaxies at similar redshifts are 
Ly$\alpha$ emitters with a restframe EW larger than 20~\AA \ and 10~\AA \ 
respectively (Shapley et al. 2003). The median restframe EW of the Ly$\alpha$ line 
for Lyman-Break galaxies (LBGs) is $\sim$0~\AA \ (about half of the LBGs 
have Ly$\alpha$ in absorption). 
The restframe EW of the Ly$\alpha$ emission line from the GRB~021004 host is 
constrained to be higher than 50~\AA \ (M\o ller et al. 2002). The restframe EW
for the host galaxy of GRB~971214 is measured spectroscopically to
be 14~\AA \ (Ahn, private communication), whereas it is 
unknown for GRB~030323 as its host is still undetected. If we 
conservatively assume that the restframe EW is above 10~\AA \ for three hosts
(971214, 011211, 030323) and above 20~\AA \ for two hosts (000926, 021004) then 
GRB host galaxies are inconsistent with being drawn randomly from the same 
Ly$\alpha$ EW distribution as the LBGs at the 
$1-0.33^3\times0.25^2 \approx 99.8$\% level.

This remarkable fact can be explained by a preference for GRB progenitors to 
be metal-poor. Ly$\alpha$ emission with EW larger than 20\AA \ is 
locally only found in starforming galaxies with [O/H] $\lesssim -0.5$  
(Charlot \& Fall 1993, their Fig.~8; Kunth et al. 1998; Kudritzki et 
al. 2000). Furthermore, Shapley et al. (2003) find that the collisionally
excited UV nebular emission lines of \ion{C}{iii}] and [\ion{O}{iii}] are
stronger than average for the quartile of the their sample with Ly$\alpha$ 
EW $>$ 20~\AA. By analogy with local starbursts this also implies low 
metallicity (Heckman et al. 1998). In the collapsar model (Woosley 1993) a 
strong stellar wind, which 
is the consequence of a high metallicity, makes it difficult to produce a GRB 
due to mass loss and loss of angular momentum (MacFadyen \& Woosley 1999). 
Therefore, a preference for GRB hosts to be metal poor is a clear prediction of
the collapsar model. Alternatively, the explanation could be an optical 
afterglow selection bias against dusty hosts. For 60--70\% of the searches for 
optical afterglows since 1997 no detection was made -- the dark burst problem
(Fynbo et al. 2001; Berger et al. 2002). Thus, the bursts for which a bright 
optical afterglow is detected, including all the bursts with detected 
Ly$\alpha$ emission from their hosts, are biased against very 
dusty host galaxies (Fynbo et al. 2001; Lazzati et al. 2002; Ramirez-Ruiz 
et al. 2002). This is important as even small amounts of dust will 
preferentially destroy Ly$\alpha$ photons due to resonant scattering (e.g. 
Ferland \& Netzer 1979). However, it remains to be shown that the majority of 
dark bursts indeed are dust obscured. In fact, several bursts have been found 
to be optically dim without significant extinction (Hjorth at al. 2002; 
Berger et al. 2002; Fox et al. 2003; Hjorth et al. 2003). The dark bursts are 
also generally fainter in X-rays (De Pasquale et al. 2003) again implying that 
they are intrinsically dim or very distant. Finally, the fraction of 
Ly$\alpha$ emitters could be larger at the faint end of the high-$z$ luminosity 
function, where most GRB hosts are found, than the fraction found for the 
bright LBGs. Shapley et al. (2003) find that among the LBGs with 
Ly$\alpha$ EW$>$20~\AA \ the EWs are largest for the faintest
galaxies, but argue that a constant fraction of Ly$\alpha$ emitters down to 
$\textrm{R}=25.5$ is consistent with the data when selection effects are taken 
into account. Furthermore, a higher fraction of Ly$\alpha$ emitters at the faint
end of the luminosity function would also imply a lower metallicity and this
is therefore not in conflict with a low metallicity preference for GRB hosts.
In conclusion, a lower metallicity of GRB hosts compared to LBGs in general 
seems to be well established. 

\section*{Acknowledgments}
We thank our anonymous referee for a very constructive report that helped
us improve the paper on several important points. We also thank Stan Woosley 
for helpful comments and the La Silla staff for excellent
support during our run. JPUF acknowledges support from the Carlsberg 
Foundation. PJ acknowledges support from The Icelandic Research Fund for 
Graduate Students, and from a special grant from the Icelandic Research 
Council. STH acknowledges support from the NASA LTSA grant NAG5--9364. 
We acknowledge benefits from collaboration within the EU FP5 Research 
Training Network ``Gamma-Ray Bursts: An Enigma and a Tool''.  
This work is supported by the Danish Natural Science Research 
Council (SNF).

\end{document}